\documentclass[prl,aps,twocolumn,superscriptaddress]{revtex4-1}
\usepackage{latexsym,amssymb,amsfonts,amsmath,amstext,graphicx,bbm,bm,hyperref,relsize,dsfont,theorem,times,lipsum}

\hypersetup{
	colorlinks=true,
	linkcolor=blue,
	filecolor=magenta,      
	urlcolor=cyan,
}

\usepackage[dvipsnames]{xcolor}

{\catcode`\|=\active
  \gdef\Braket#1{\begingroup
\mathcode`\|32768\let|\BraVert\left<{#1}\right>\endgroup}}
\def\BraVert{\egroup\,\mid\,\bgroup}

\begin{document}

\title{Strong Thermo-mechanical Squeezing in a far detuned Membrane-in-the-middle System}

\author{Sameer Sonar}
\email{sameersonar@iitb.ac.in}
\affiliation{Department of Physics, Indian Institute of Technology-Bombay, Powai, Mumbai 400076, India.}
\affiliation{Huygens-Kamerlingh Onnes Laboratorium, Universiteit Leiden, 2333 CA Leiden, The Netherlands}

\author{Vitaly Fedoseev}
\affiliation{Huygens-Kamerlingh Onnes Laboratorium, Universiteit Leiden, 2333 CA Leiden, The Netherlands}

\author{Matthew J. Weaver}
\affiliation{Department of Physics, University of California, Santa Barbara, USA}

\author{Fernando Luna}
\affiliation{Department of Physics, University of California, Santa Barbara, USA}

\author{Elger Vlieg}
\affiliation{Huygens-Kamerlingh Onnes Laboratorium, Universiteit Leiden, 2333 CA Leiden, The Netherlands}

\author{Harmen van der Meer}
\affiliation{Huygens-Kamerlingh Onnes Laboratorium, Universiteit Leiden, 2333 CA Leiden, The Netherlands}

\author{Dirk Bouwmeester}
\affiliation{Department of Physics, University of California, Santa Barbara, USA}
\affiliation{Huygens-Kamerlingh Onnes Laboratorium, Universiteit Leiden, 2333 CA Leiden, The Netherlands}

\author{Wolfgang L\"{o}ffler}
\email{loeffler@physics.leidenuniv.nl}
\affiliation{Huygens-Kamerlingh Onnes Laboratorium, Universiteit Leiden, 2333 CA Leiden, The Netherlands}

\date{\today}

\begin{abstract}
We demonstrate 8.5 dB thermal squeezing of a membrane oscillator using the dynamical backaction effect and electrostatic feedback in an optomechanical membrane-in-the-middle setup. We show that strong squeezing can be obtained even in the far detuning regime of a sideband-resolved system. By using the dielectrophoretic force of a metallic needle kept in close proximity to the membrane, we implement the one-quadrature active feedback scheme to prevent the divergence of the amplified quadrature and surpass the 3 dB limit of mechanical squeezing. We also discuss different regions of the sideband spectrum where strong squeezing can be obtained. Although the demonstration here is classical, this technique is equally applicable to prepare the mechanical oscillator in a quantum squeezed state.
\end{abstract} 
\maketitle

\makeatletter

Methods to prepare a mechanical oscillator in a squeezed state \cite{rugar1991} are important to enhance the read-out sensitivity and to reduce the measurement backaction in the quadrature of interest. We discuss here quadrature-squeezed states where the noise in the $\sin\omega t$ and $\cos\omega t$ quadratures are not equal, in contrast to coherent and thermal states. Several methods have been proposed to generate squeezed mechanical states both in the classical \cite{rugar1991,difilippo1992classical} and the quantum regime \cite{braginsky1980quantum,clerk2008back,vanner2011pulsed,szorkovszky2011mechanical,you2017strong}. 
In the quantum regime, quantum backaction evading measurements have demonstrated noise reduction close to the zero-point motion \citep{hertzberg2010back} and even beyond \citep{suh2014mechanically,wollman2015quantum, lecocq2015quantum,pirkkalainen2015squeezing, lei2016quantum}.

Classical squeezed mechanical states, or thermally squeezed states have been generated in opto or electro mechanical systems by applying a parametric force \citep{rugar1991,Briant2003,vinante2013feedback,szorkovszky2013strong,poot2014classical,poot2015deep}, by optically modulating the spring constant \citep{pontin2014squeezing}, by fast switching between two trapped frequencies \citep{rashid2016experimental} and by quantum non-demolition measurements \cite{vanner2013cooling}. Further, it is also possible to generate squeezed states involving a quadrature of two different mechanical oscillators \citep{mahboob2014two,patil2015thermomechanical,pontin2016dynamical,
ockeloen2018stabilized}. 
 

The thermal excitation of a mechanical oscillator leads to a Gaussian noise distribution, where the noise amplitudes in both quadratures are equal. Parametric modulation of the spring constant at twice the mechanical frequency breaks this degeneracy of the two quadratures \cite{rugar1991}. The thermal noise in the quadrature which is in-phase with this modulation increases while the thermal noise in the orthogonal out-of-phase quadrature decreases. When the noise amplitude in one quadrature is reduced by a factor of 3 dB, the orthogonal one becomes infinitely large.
Above the 3 dB limit, the system behavior becomes chaotic due to the parametric instability in the diverging quadrature \cite{Briant2003}. Fortunately, it is possible to prevent the divergence and the parametric instability and enable squeezing of the orthogonal quadrature beyond the 3 dB limit. Experiments performed to surpass this limit include active feedback on the diverging quadrature \cite{vinante2013feedback,poot2015deep}, real time squeezing phase adjustment \cite{poot2014classical}, detuned pump \cite{szorkovszky2013strong} and reservoir engineering \cite{lei2016quantum}.

The radiation force acting on a mechanical oscillator in an optomechanical system has the important property that its response to a change in the oscillator position is delayed by approximately the cavity decay time. This delay modifies the damping and the frequency of the mechanical oscillator as a function of the laser detuning and is called the dynamical backaction effect \cite{sheard2004observation}. In a sideband unresolved system where the mechanical frequency $\Omega_m$ is smaller than the cavity decay rate $\kappa$, the dynamical backaction is significant only if the laser is near-resonance with the optical cavity mode. To our knowledge, this is the only regime of the sideband spectrum in which optomechanical squeezing beyond 3 dB has been demonstrated \cite{pontin2014squeezing}. On the other hand, in the sideband resolved regime ($\Omega_m > \kappa$), side-band cooling can be used to cool the mechanical oscillator to its quantum ground state \cite{Aspelmeyer2014}. Also, using a far-detuned laser is important in experiments where one wants to perform a squeezing operation on an arbitrary initial state without any additional heating or cooling effects. 

Here, we show optomechanical squeezing for a well sideband-resolved optomechanical system with $\Omega_m/\kappa\approx9$, where the dynamical backaction is efficient even in the far detuned regime. We obtain 8.5 dB mechanical squeezing using the dynamical backaction effect at 1.43$\times\Omega_m$ laser detuning. This strong squeezing despite large detuning is also facilitated by using a mechanical oscillator with a reasonably high Q factor of $3\times 10^5$ at room temperature that enables higher spring constant modulation \cite{vinante2013feedback} and consequently more squeezing.

We first present our membrane-in-the-middle setup and discuss parametric squeezing for different laser detunings. Then we show 3 dB squeezing accompanied by the parametric instability in one quadrature. Finally, we demonstrate how to surpass the 3 dB limit by applying feedback onto the diverging quadrature of the mechanical oscillator.

\begin{figure}[t]
\includegraphics[width=\linewidth]{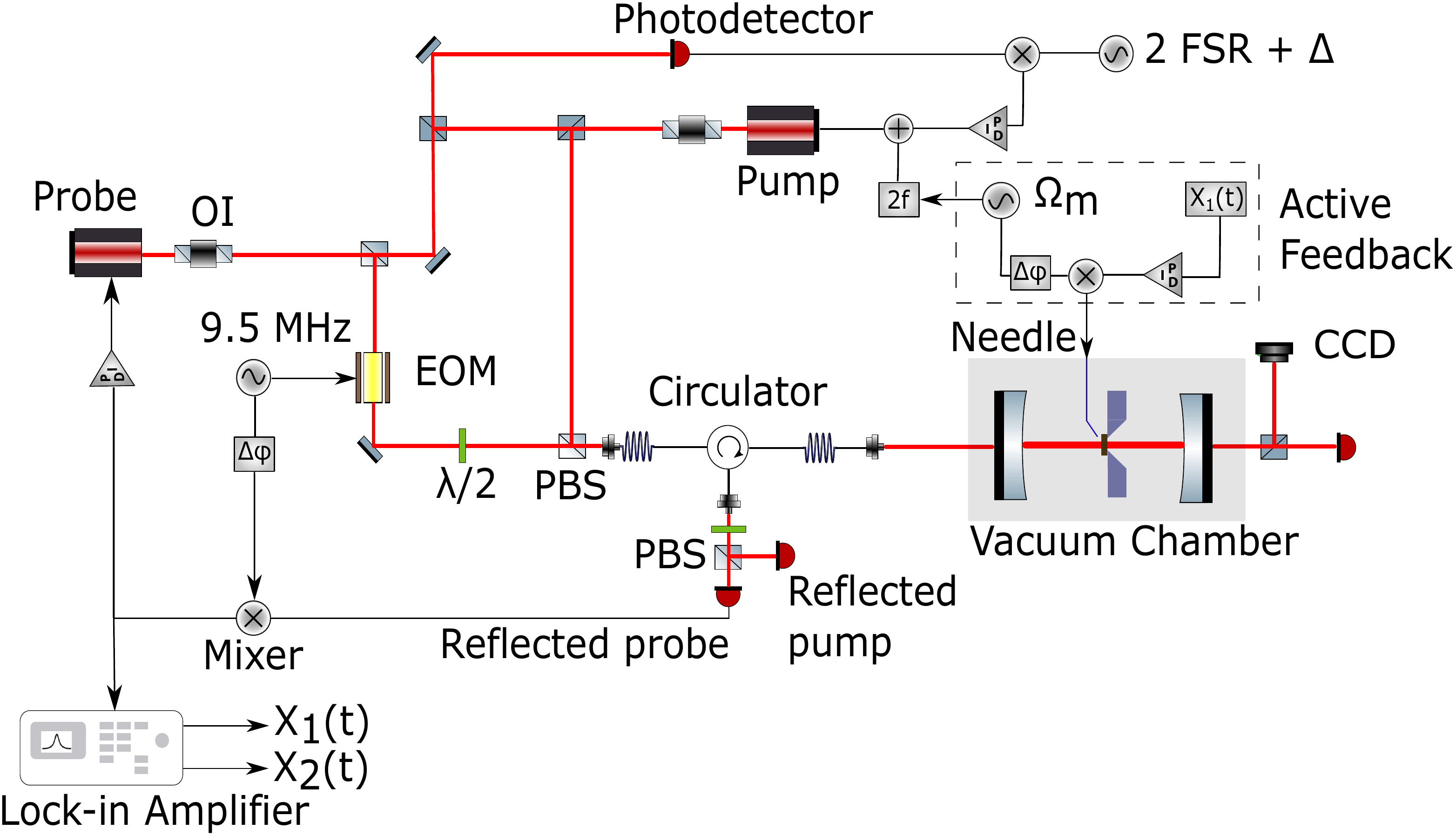}
\caption{Experimental membrane-in-the-middle setup. The probe laser is locked to the optical cavity resonance frequency with the Pound-Drever-Hall (PDH) technique and used to record the motion of the membrane, the pump laser applies the parametric force, and the needle generates the parametric electrophoretic force. EOM: electro optical
modulator, OI: optical isolator, PID: proportional-integral differential feedback controller, PBS: polarizing beam splitter, CCD: charge coupled device.}
\label{optical_setup}
\end{figure}


Experimental Setup: Our optomechanical system consists of a 98 mm Fabry-Perot cavity operating at 1064 nm. High quality thin film mirrors consisting of alternating Ta$_2$O$_5$/SiO$_2$ layers are mounted at the end of a 98 mm Invar spacer tube similar to \cite{jayich2008dispersive}. The radius of curvature of the mirrors is 5 cm which gives strong near-critical focusing at the center of the cavity. The setup is placed inside a vacuum chamber at a pressure of $10^{-6}$ mbar.

The mechanical oscillator used in this scheme is a high stress 50 nm silicon nitride semi-transparent membrane manufactured by NORCADA-Inc. The fundamental mode has a resonance frequency $\Omega_m/2\pi\approx385$ kHz, effective mass $m_{\text{eff}}\approx33$ ng and a Q factor of $3\times 10^5$ at room temperature. The membrane is placed in the middle of the cavity, and its tilt and axial position are controlled using 3 piezo motors. The axial position of the membrane with respect to the standing wave pattern inside the cavity is chosen to give large first order single photon optomechanical coupling $g_0$ \cite{thompson2008strong}. We obtain an optical finesse of the fundamental (Gaussian) cavity mode of 33000, and a membrane coupling of $g_0/2\pi=3$ Hz from calculating the derivative of the dispersion curve.

The optical setup is shown in Fig.~\ref{optical_setup}. We use two lasers: The probe laser ($\approx$ 10 $\mu$W) is used for sensitive readout of the motion of the membrane. We use the Pound-Drever-Hall (PDH) method \citep{pdh} to lock the laser to the cavity. The PDH error signal is demodulated at the mechanical frequency and is used to observe the real time phase space trajectory of the fundamental mechanical mode: $x(t)=X_1(t)\sin(\Omega_mt)+X_2(t)\cos(\Omega_mt)$. A second more powerful ($\approx$ 400 $\mu$W) pump laser is used to impart radiation force onto the membrane. This laser is locked to the probe laser with an optical phase locked loop and a 2 FSR offset to avoid any unwanted interference. Using an offset of 2 FSR is advantageous over 1 FSR because unavoidable nm-scale drifts of the membrane position lead (ideally) to no changes in the frequency spacing of cavity resonances separated by 2 FSR. \citep{jayich2008dispersive}. The other means to influence the motion of the membrane is the dielectrophoretic force of a metal needle situated near the membrane \cite{buters2017high}. The needle is far enough from the centre of the membrane that it does not interfere with the optical beam but is close enough to create a large electric field gradient.

\begin{figure}[t]
\includegraphics[width=\linewidth]{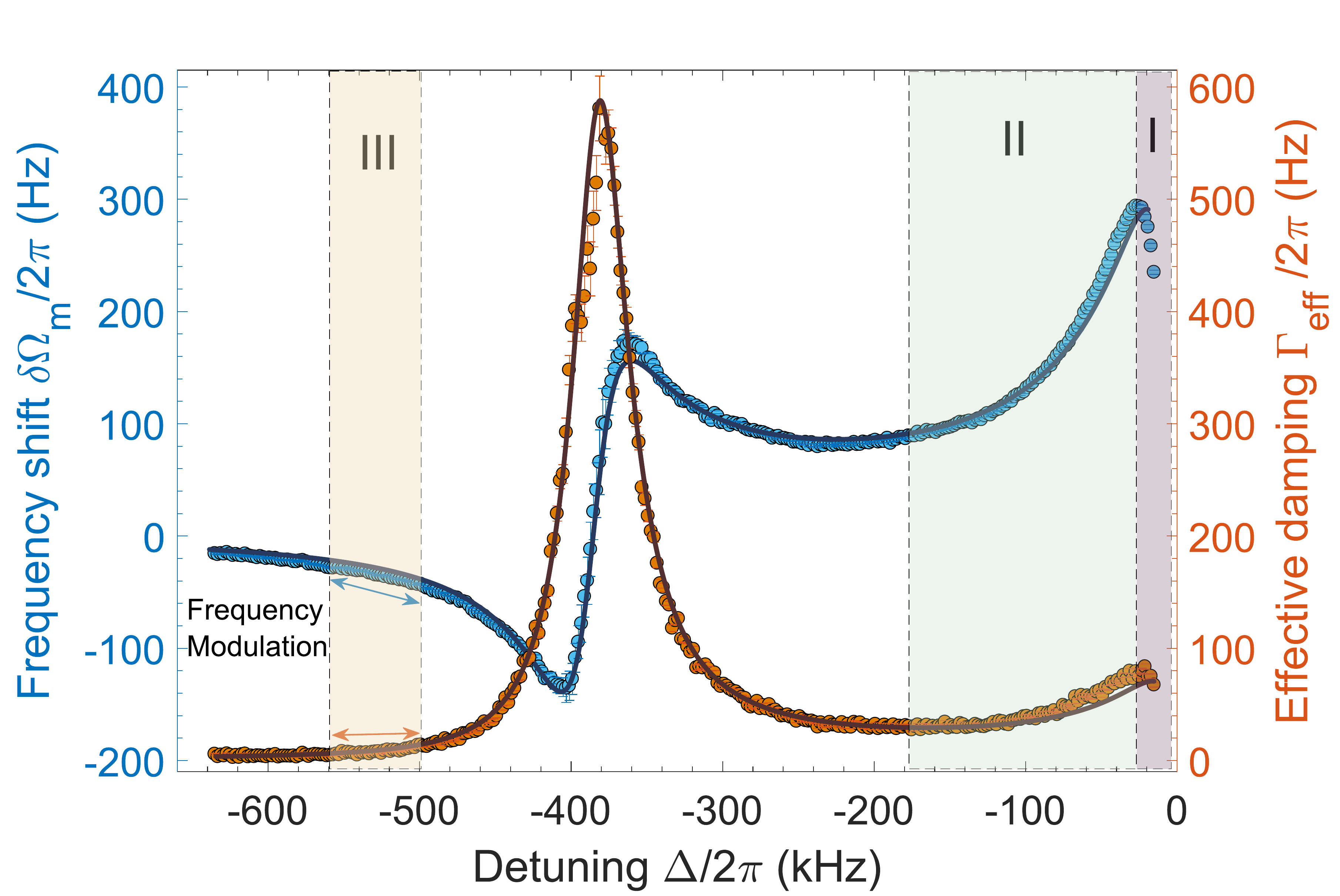}
\caption{Measurement of the effective damping $\Gamma_{\text{eff}}$ (red) and mechanical frequency shift $\delta\Omega_{\text{m}}$ (blue) depending on the pump laser detuning from the cavity resonance (at $\Delta=0$). Region I, II and III show potential regions to perform spring constant modulation by frequency-modulation of the laser. The solid lines represent theoretical fits to the experimental data.}
\label{detuning_sweep}
\end{figure}

We follow the analysis given in \citep{vinante2013feedback} for parametric driving, taking into account the additional optical damping $\Gamma_{\text{opt}}$. The equation for a squeezed thermo-mechanical oscillator is given by
\begin{equation}
\ddot{x}+\Gamma_{\text{eff}}\dot{x}+\Omega_m^2\left[1-\frac{2g}{Q^\prime}\sin(2\Omega_mt+\phi)\right]x=F_{\text{th}}(t).
\label{squeezed_oscillator_equation}
\end{equation}

Here $\Gamma_{\text{eff}}=\Gamma_m+\Gamma_{\text{opt}}$ is the effective damping, $Q^\prime=\Omega_m/\Gamma_{\text{eff}}$ is the effective Q factor, $F_{\text{th}}(t)$ is the thermal Langevin force which is responsible for the Brownian motion and $g$ is the gain parameter of the spring constant modulation. Since this modulation is exactly at twice the mechanical frequency, the two quadratures experience different forces. The one which is in phase with this modulation is amplified in variance and the orthogonal one is squeezed. For $\phi=0$, the variance of each quadrature, $X_1$ and $X_2$, is given by
\begin{equation}
\langle X_1^2\rangle=\frac{k_BT_{\text{eff}}}{m_{\text{eff}}\Omega_m^2}\frac{1}{(1-g)},\langle X_2^2\rangle=\frac{k_BT_{\text{eff}}}{m_{\text{eff}}\Omega_m^2}\frac{1}{(1+g)}.
\label{variances_squeezing}
\end{equation}
Here $k_B$ is the Boltzmann constant and $T_{\text{eff}}$ is the effective temperature of the fundamental mode which takes into account the optical damping. Changing the modulation phase by $\phi$ changes the squeezing direction by $\phi/2$ \cite{poot2014classical}. We change this offset to align the squeezing axis with the $X_2$ quadrature. Fig.~\ref{detuning_sweep} shows the measurement of the effective damping $\Gamma_{\text{eff}}$ and the mechanical frequency shift $\delta\Omega_m$ as a function of pump laser detuning $\Delta$. We place a laser beam in the shaded region III which is $1.43\times\Omega_m$ detuned on the red side and perform frequency modulation. The non-zero slope of the frequency shift curve leads to modulation of the spring constant. We note that the frequency spectrum of our membrane is slightly anharmonic and the higher order modes deviate slightly from their expected values by few hundreds of Hertz, such that the frequency of the (2,2) mode is not exactly twice that of the (1,1) mode. This deviation works in our favor because driving the membrane at twice the mechanical frequency of the (1,1) mode does not excite the (2,2) mode. The feedback bandwidth of the optical phase locked loop is much smaller than $\Omega_m$ and hence the frequency modulation at $2\Omega_m$ does not affect the optical phase locked loop. In the shaded region, the change in frequency shift is 2.7 times higher than the change in effective damping and we treat the effective damping as a constant to simplify the analysis. This approximation is not valid, for example, near the red sideband where the effective damping is highly nonlinear and completely overwhelms the frequency shift. In this case, one also needs to consider the modulation of the effective damping in Eq.~\ref{squeezed_oscillator_equation} which leads to cross coupling between the two quadratures and the variances in the two quadratures cannot be written in a decoupled way as in Eq.~\ref{variances_squeezing}.

Fig.~\ref{detuning_sweep} also shows other regions of the sideband spectrum where the derivative of the frequency shift curve $\delta\Omega_{\text{m}}$ is higher than that of the effective damping $\Gamma_{\text{eff}}$ and where strong squeezing can be obtained. For instance, the study in \citep{pontin2014squeezing} operates in region I close to zero detuning. Our sideband resolved system is unstable in this region because the mechanical oscillator gets driven as soon as the pump beam is slightly detuned on the blue side, making this region impractical to use. Additionally, region I and II have a larger second order derivative of the frequency shift and effective damping curves making the spring constant modulation in Eq.~\ref{squeezed_oscillator_equation} `non-linear'. We select region III to maximally decouple optical damping from squeezing.


\begin{figure}[t]
\includegraphics[width=\linewidth]
{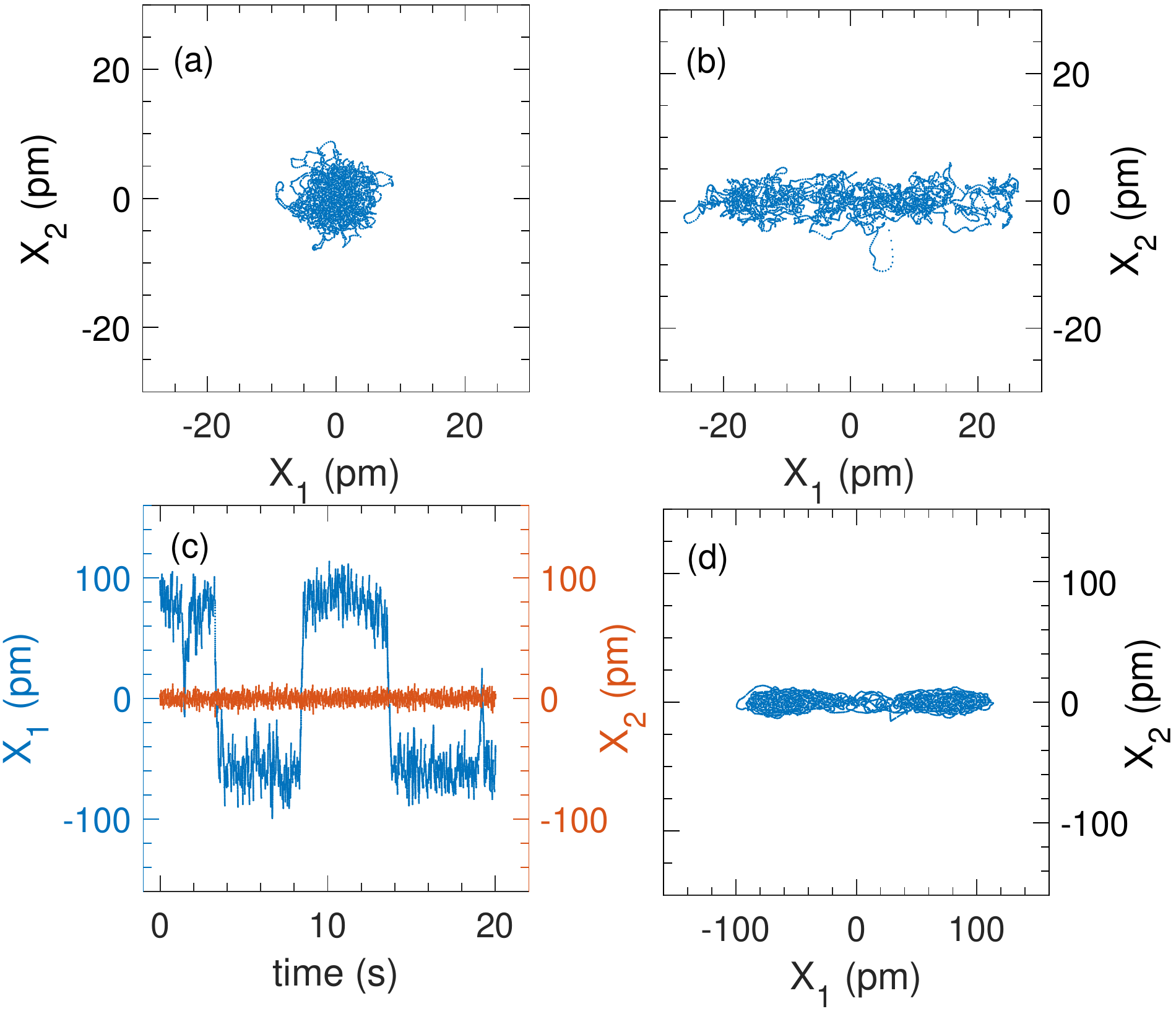}
\caption{(a) Phase space trajectory of the mechanical oscillator with an effective temperature $T_{\text{eff}}=120$ K. (b) Squeezed trajectory with gain parameter $g=0.96$ after aligning the squeezing direction with $X_2$ quadrature. (c) Bistability in the $X_1$ quadrature and squeezed distribution in the $X_2$ quadrature for $g > 1$ as a function of time (c) and in phase space (d).}
\label{phase_plot}
\end{figure}

Figure \ref{phase_plot}(a) shows the phase space thermal distribution of the resonator for the case when the pump laser is at $\Delta=1.43\times\Omega_m$ on the red side (in region III). The resulting optical damping leads to an effective temperature $T_{\text{eff}}=120$ K. The data is recorded for 5 seconds with a demodulation bandwidth of 138 Hz. The distribution is Gaussian in both quadratures with equal variances. Applying $2\Omega_m$ frequency modulation to the laser leads to a squeezed distribution as shown in Fig.~\ref{phase_plot}(b). Increasing the gain parameter $g$ slightly above 1 leads to the parametric instability whereby the thermal distribution gets shifted from the origin. The amplified $X_1$ quadrature shows bistability because the equation of motion is invariant under a $\pi$ phase shift \cite{Briant2003} while the distribution in $X_2$ is still squeezed. Fig.~\ref{phase_plot}(c), (d) show this behavior as a function of time and in phase space respectively.   

Parametric feedback: The 3 dB limit imposed by the parametric squeezing is not a fundamental limit and can be surpassed by preventing the divergence of the amplified quadrature. We use the one-quadrature active feedback technique \cite{vinante2013feedback} to sense the motion in the amplified quadrature and apply a force to counteract that motion. For this purpose, we use a sharp tip metal needle close to the membrane to generate a dielectrophoretic force. The feedback force added to Eq.~\ref{squeezed_oscillator_equation} and the modified variances in the two quadrature are now given by
\begin{align}
F_{\text{fb}}(t)&=-h\frac{\Omega_m^2}{Q^\prime}X_1(t)\cos(\Omega_mt),\\
\langle X_1^2\rangle&=\frac{k_BT_{\text{eff}}}{m_{\text{eff}}\Omega_m^2}\frac{1}{(1-g+h)},\langle X_2^2\rangle=\frac{k_BT_{\text{eff}}}{m_{\text{eff}}\Omega_m^2}\frac{1}{(1+g)},
\label{squeezing_variances_feedback}
\end{align}
where $h$ is the dimensionless feedback parameter. The feedback circuit is shown in the dashed box in Fig.~\ref{optical_setup}. We mix the selected quadrature ($X_1$ here), modified by appropriate PID parameters, with $\cos\Omega_m t$ and send it to the needle. The squeezing gain parameter $g$ of $X_1$ quadrature is shifted by $h$ while the orthogonal quadrature $X_2$ remains independent and can now be squeezed beyond 3 dB because the $X_1$ quadrature does not diverge for $g\rightarrow$ 1.

\begin{figure}[t]
\includegraphics[width=\linewidth]
{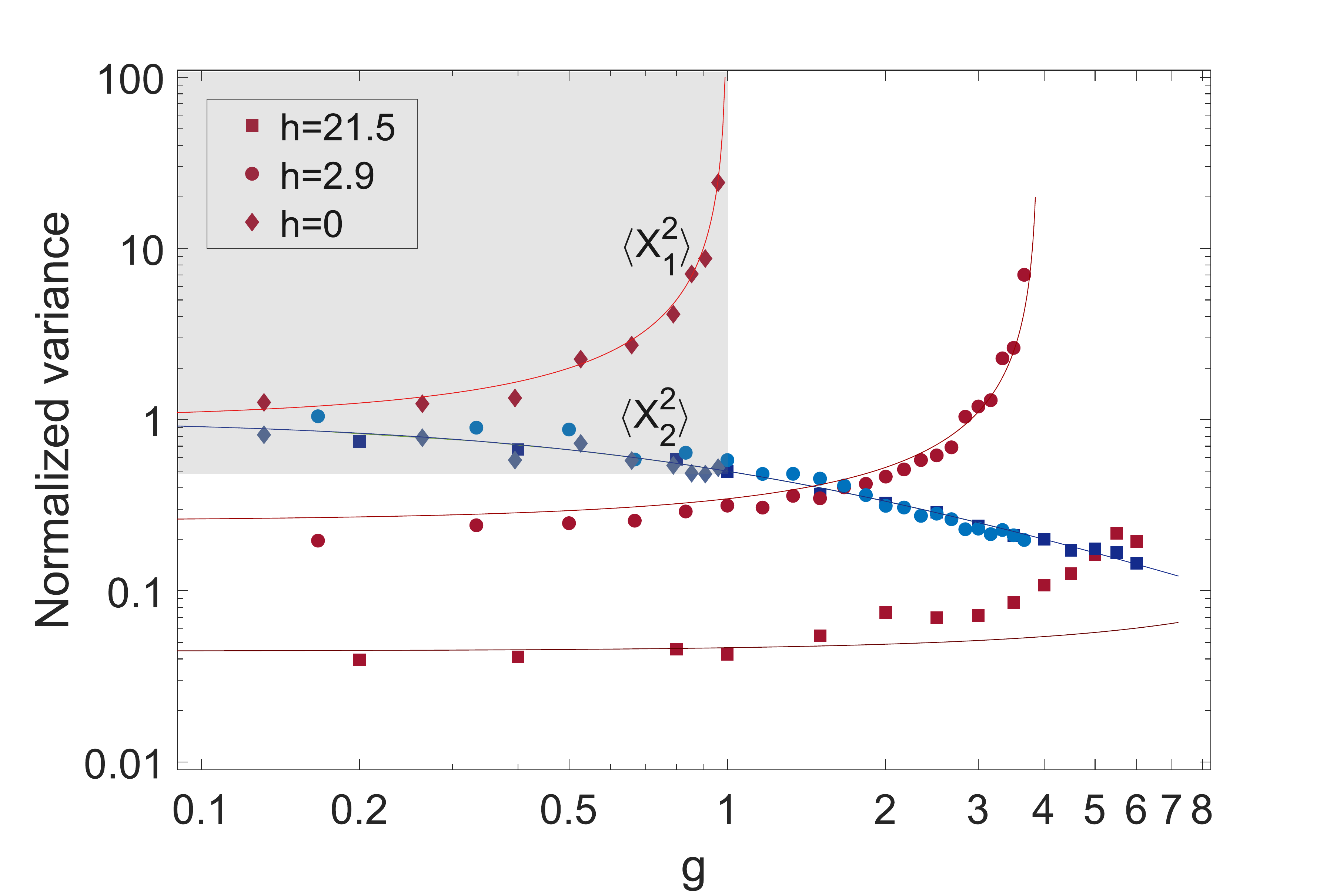}
\caption{Normalized variances in $X_1$ and $X_2$ quadratures as a function of the gain parameter $g$. The shaded area highlights the allowed parameter space without any feedback. Diamonds: $h=0$, circles: $h=2.9$ and a maximum of 7.03 dB squeezing, squares: $h=21.5$ and a maximum of 8.5 dB squeezing. $T_{\text{eff}}$ is 120 K for $h=0$ \& 2.9 respectively and 182 K for $h=21.5$. Solid lines represent theoretical fits based on Eq.~\ref{squeezing_variances_feedback}.}
\label{plot_variance}
\end{figure}

Figure~\ref{plot_variance} shows the quadrature variances for different strengths of active feedback. We normalize the variances with respect to the thermal distribution when the pump beam is 1.43$\times\Omega_m$ detuned on the red side. This far detuning ensures minimal sideband cooling of the initial state and allows us to perform a squeezing operation largely independent of sideband cooling. The shaded area shows the allowed parameter space without any active feedback. Increasing the feedback parameter to $h=2.9$ shifts the diverging curve (red diamonds) to the right and a maximum of 7.03 dB squeezing is obtained. For $h=21.5$, a lower pump power was used which resulted in an effective temperature of 182 K. A maximum squeezing of 8.5 dB is obtained in this case. At large gain parameter $g$, the variance in $X_1$ diverges from the theoretical fit; we suspect that we reach the non-linear region of the frequency shift and the effective damping curves outside of the region III in Fig.~\ref{detuning_sweep}. Additionally, there could be a slight mismatch between the feedback direction and the diverging quadrature \cite{poot2015deep}.

Discussion: In this study, we have shown that strong thermally squeezed states can be obtained using the dynamical backaction effect even in the far detuned regime. We have implemented a simple electrical feedback scheme to surpass the 3 dB limit and confine the motion of the mechanical oscillator in a strongly squeezed state. By selecting the far detuned region, we are able to minimize optomechanical cooling so that our squeezing operation can be performed on an arbitrary initial state. This method is effective to perform many quantum experiments with squeezed states such as transfer between different mechanical oscillators \citep{weaver2017coherent}, electro-optomechanical transduction \citep{mcgee2013mechanical}, possible studies of decoherence in macroscopic objects \citep{marshall2003towards,weaverphonon} and the enhancement of quantum synchronization \cite{sonar2018squeezing}. Having a resolved sideband system has the advantage that there are several regions of the sideband spectrum where strong spring constant modulation can be obtained. However, the optical readout of the mechanical motion is also delayed due to the high finesse cavity, which limits the amount of feedback cooling one can get in one quadrature. 

To sum up, there are three factors that are currently limiting the amount of squeezing: the strength of the one- quadrature active feedback, the strength of the spring constant modulation and the quality factor of the mechanical oscillator. The second issue can in principle be solved by using two driving tones, one on the red sideband and the other on the blue sideband and frequency modulating both of them at twice the mechanical frequency. In this way the damping cancels out whereas the frequency shifts due to the two tones add up giving a large spring constant modulation. This method can also be used in the reversed dissipation regime \citep{nunnenkamp2014quantum}, where the role of the mechanical oscillator and the optical oscillator are interchanged, to obtain strong squeezed light. Lastly, by using membranes with exceptionally high quality factor such as in \citep{tsaturyan2017ultracoherent}, combined with our sideband resolved optical cavities, it should be possible to prepare the mechanical resonator in a truly quantum squeezed state. During preparation of this manuscript, the key requirement of optical cooling of a membrane oscillator close to the quantum ground state has been achieved \citep{rossi2018b} by using optical feedback with a $Q=10^9$ soft-clamped membrane in a cryostat with a base temperature of $T\sim 10$ K.  \\

{\bf Acknowledgements --} This work is part of the research program of the Foundation for Fundamental Research (FOM)
and of the NWO VICI research program, which are both part of the Netherlands Organization
for Scientific Research (NWO).\\
The authors would like to thank Kier Heeck for technical assistance. The authors are also grateful for useful discussions with Frank Buters and Sai Vinjanampathy.

\bibliographystyle{naturemag}
\bibliography{biblio_WL}

\end{document}